\documentstyle[aps]{revtex}
\begin{document}
\draft
\newcommand{\be}{\begin{equation}}
\newcommand{\ee}{\end{equation}}
\newcommand{\bea}{\begin{eqnarray}}
\newcommand{\eea}{\end{eqnarray}}

\title{Effect of Boundary Conditions on Fluctuations Measures
\thanks{journal ref: Phys. Rev.~E, VOLUME 57, PAGE R3699, 1998}}
\author{Debabrata Biswas \thanks{email:~dbiswas@apsara.barc.ernet.in}}
\address{Theoretical Physics Division \\
Bhabha Atomic Research Centre \\
Trombay, Mumbai 400~085}

\date{\today}
\maketitle
\begin{abstract}
A change in boundary conditions (BC) from uniform Dirichlet
to non-identical BC on the edges of a triangular
billiard often brings
about a dramatic change in quantum spectral fluctuations.
We provide a theory for this based on periodic orbits
and show that non-identical BC on adjacent edges can lead
to a {\em quantum splitting} (QS) of periodic orbit families which
results in a significant change in the form factor. Thus, the
classical spectrum alone cannot determine quantum correlations.

\end{abstract}
\vskip 0.4 in
\pacs{PACS numbers: 05.45.+b, 03.65.Sq}

\par
Recent studies \cite{AAA,BK} on fluctuation measures in chaotic systems
seem to indicate that the quantum correlations are
fully determined by the classical spectrum of the
Perron-Frobenius operator \cite{CE}. Using different approaches,
Agam, Altshuler and Andreev \cite{AAA} and Bogomolny and Keating
\cite{BK} show that the diagonal and off-diagonal part
of the density-density correlation are related to a purely classical
quantity, which under some approximation reduces to the
classical zeta function \cite{CE}.

\par
There are several fallouts of the AAA-BK theory. One that has been
scrutinized recently by Prange \cite{Prange} concerns possible
deviations from random matrix theory results and the conditions
under which this can be observed. Another consequence (and
one that is of relevance here) is
the absence of parity effects in fluctuation measures. In other
words, the AAA-BK theory predicts that quantum systems having
the same classical dynamics exhibit identical quantum
correlations. There is however a  tacit assumption in the
BK approach which leads us to this conclusion - that degenerate
periodic orbits with identical actions and stabilities also
have the same quantum phase. There can be instances however
when this is not true. For example, arithmetic billiards
abound in degenerate periodic orbits and exhibit Poisson
fluctuations when the boundary conditions are Dirichlet
\cite{bertie}. However, when the boundary conditions  
are not identically Dirichlet, pairs of degenerate periodic
orbits can have phases differing by $\pi$ leading
to a net decrease in the form factor \cite{steiner95}.

\par The effect of boundary conditions can be even
more significant in planar triangular billiards and we shall
deal with these henceforth. Of these, the ones that
are integrable have internal angles of the form $\pi/n_i$ and
in all these cases, degeneracies exist in the classical
periodic orbit actions of topologically distinct orbits
leading to non-universal spectral fluctuations \cite{dutch,see_bis91}.
Thus, the spectral rigidity \cite
{mehta}, $\Delta_3(L)$, increases with a slope larger than
${1\over 15}$ \cite{d3_integrable} in the
region $L << L_{max}$ where $L_{max}$ is determined
by the frequency of slowest oscillation in the quantum
density $\rho(E)$ \cite{ber85}.

\par
Generic rational triangles on the other hand have internal
angles of the form $\pi m_i/n_i$ where
$\Pi_{i=1}^{3} m_i \neq 1$. These are referred to as
pseudo-integrable (PI) billiards \cite{RB,fnote_rational}.
As in integrable systems,
their invariant surface in phase space is 2-dimensional
but the topology is that of a
sphere with multiple holes and not a torus \cite{intro_genus}.
This difference
leads to a rather dramatic change in the quantum eigenstates.
The eigenfunctions for example often exhibit irregular nodal
patterns and a Gaussian amplitude distribution \cite{db1}
while fluctuation measures display a behaviour \cite{dutch,shudo}
that ranges from the integrable \cite{dutch} to the 
chaotic \cite{berwil,cheon,db_irr,dutch}
limits.

\par
There are several interesting questions concerning quantum
fluctuations that polygonal billiards throw up. 
A point that has often been debated
is the role of diffractive periodic orbits in determining
spectral measures \cite{SHSH}. Admittedly, the quantum spectrum does
know about these orbits \cite{pavloff}
though its importance in determining
spectral measures is possibly negligible \cite{db_rapid}.
A related question concerns the effect of boundary conditions on
spectral fluctuations. To illustrate this, we refer to
fig. 1 where the rigidity, $\Delta_3$ is plotted
as a functions of $L$ for the right triangle $(\pi/2,\pi/3,\pi/6)$
with (a) Dirichlet boundary condition on all the three edges
and (b) Neumann boundary condition on the edges enclosing the
right angle and Dirichlet BC on the third. One might argue
that the crossover is related to the fact that case (a)
is integrable while (b) is not \cite{fnote_rhombus}. 
We have thus verified that there is indeed a shift with
boundary conditions in genuine pseudo-integrable enclosures
such as the $(\pi/2,3\pi/10,\pi/5)$  triangle.

\par At first glance, it may seem that apart from an
overall phase
factor, the contribution of each periodic orbit family
is identical for the two cases of the $(\pi/2,\pi/3,\pi/6)$
triangle. 
We shall demonstrate here that this is not the case. To this
end, consider the family of periodic orbits shown in
fig. 2. Orbits F1 and F2, belong to the same family {\em classically} 
though the quantum phase accumulated by these differ
by $\pi$ when edges 1 and 3 have Neumann boundary
conditions (we refer to this as case (b) while (a) denotes
Dirichlet BC on all edges). In other words, the family
of periodic orbits
split up in the {\em quantum-mechanical} sense in case (b)
while case (a) preserves the full classical family.
The semiclassical density of states,

\be
\rho(E) \simeq \rho_{av}(E) + \sqrt{{1\over
8\pi^3}} \sum_p \sum_{r=1}^\infty {a_p\over
\sqrt{krl_p}} \cos(krl_p - {\pi\over 4} - rn_p\pi) \label{eq:basic}
\ee

\noindent
for the two cases are thus distinct. In the above, $E = k^2$ while
$a_p$ refers to the
area occupied by a primitive periodic orbit family
characterized by their length $l_p$ and {\em the
number, $n_p$ of reflections from Dirichlet edges}.
Eq. (\ref{eq:basic}) has only the leading order fluctuation in the
density of states. It neglects the contribution of isolated
periodic orbits and diffractive periodic orbits. Also note that orbits
that occur in families necessarily undergo even number of reflections
from the edges so that the net phase in case (a) is zero.

\par With this background, we are now ready to explore
the effect of BC on spectral measures. For two point
correlations such as $\Delta_3(L)$ and $\Sigma_2(L)$, the
central object is the form factor
$\phi(T) = \int_{-\infty}^{^\infty} R_2(x) \exp(ixT/\hbar) dx$
where $R_2(x)$ is the 2-point spectral correlation function
($R_2(x) = \left < \rho(E+x)\rho(E) \right >$).
Expressed in terms of periodic orbits,
$\phi(T) = \left < \sum_i\sum_j A_i A_j
\cos(S_i - S_j) \delta(T - (T_i + T_j)/2) \right >$
where $A_i = Ca_i/\sqrt{kl_i}$ for marginally unstable
billiards, $S_i = kl_i - \pi/4 - n_i\pi$, $T_i =
\partial S_i/\partial E$ and $C = \sqrt{1/(32\pi^3)}$.

\par
It is customary to analyze the diagonal and off-diagonal
part of $\phi(T)$ separately and we first show
that for case (b), the diagonal contribution
$\phi_D(T) = \left < \sum_i A_i^2 \delta(T - T_i) \right >$
is smaller as compared to case (a). Let us assume, that
the family labeled by $i$ splits up quantum mechanically
in case (b) into two parts occupying areas $a_{i1}$ and
$a_{i2}$ respectively where $a_{i1} + a_{i2} = a_i$.
Its contribution to $\phi_D(T)$ is thus proportional
to $a_{i1}^2 + a_{i2}^2$ while in case (a),
it is proportional to $a_{i1}^2 + a_{i2}^2 + 2a_{i1}a_{i2}$.
Further, since the two parts of the classical family
have a different phase in case (b), there is an off-diagonal (OD)
contribution from within this classical family. Its
magnitude is proportional to $2 a_{i1}a_{i2}\cos(\pi)$ so that
the net decrease in contribution of a single classical family
is proportional to $4a_{i1}a_{i2}$. 

\par
Note that the off-diagonal part of the form factor has
cross contributions as well where parts of two distinct
classical families are involved. When the classical
dynamics is integrable and no QS
occurs, the OD contributions average to
zero in the absence of {\em degeneracies} amongst periodic
orbit actions. Thus the diagonal contribution equals
the asymptotic value of $\phi(T)$ which equals
$\rho_{av}/2\pi$. This asymptotic law is referred to
as the semiclassical sum rule \cite{ber85}.
Even in the presence of QS,
the semiclassical sum rule holds. Thus, cross
terms involving parts of distinct classical families
do contribute in case (b).
In summary then,
the following comparison between cases (a) and (b)
can be made when there are {\em no degeneracies} in the lengths of
topologically distinct periodic orbits.
For $ T << T_H$, the form factor equals

\bea
\phi(T) & = & \left < C^2 \sum_i {a_i^2 \over kl_i} \delta( T - T_i)
\right > \ldots case~(a)  \label{eq:diag_a}   \\
& = & \left < C^2 \sum_i {\{a_i(2\alpha_i -1)\}^2
\over kl_i} \delta( T - T_i) \right > \ldots case~(b) \label{eq:diag_b}
\eea

\noindent
while in both cases, $\phi(T) = \rho_{av}(E)/(2\pi)$ as
$ T \rightarrow \infty$. Here $a_{i1} = \alpha_ia_i$, 
$a_{i2} = (1-\alpha_i)a_i$ and
$T_H$ is the Heisenberg time.

\par
Note that in an integrable enclosure, the areas $a_i$ are 
identical for almost all orbit families so that for case (a),
a straightforward application of the proliferation law 
for periodic orbit families leads to the conclusion that 
$\phi(T)$ is constant and equals $\rho_{av}(E)/ 2\pi$.
For case (b) however, the factor $(2\alpha_i -1)^2$ 
varies with the orbit and depending on the splitting mechanism,
$\phi(T)$ may be explicitly $T$ dependent even in an ``integrable''
enclosure \cite{as_in_pi}.

\par In order to concretize these notions, let us take another
look at the ($\pi/2,\pi/3,\pi/6$) enclosure of fig. 1. For this
integrable billiard, the length spectrum can be expressed in
terms of winding numbers on tori and it is easy to verify 
that there exists degeneracies in the lengths of topologically
distinct periodic orbits. For case (a) then, the sum in
Eq.(\ref{eq:diag_a}) is over distinct lengths $l_i$
instead of topologically distinct orbits. Correspondingly,
the area $a_i$ should now be interpreted as the total
area occupied by all orbit families having length $l_i$.
An immediate consequence is that $\phi(T)$ is no longer
a constant for all $T$ since the degree of degeneracy varies
with length \cite{see_bis91}. A plot of 
$I(\tau) = {2\pi\over \rho_{av}}\int \phi(\tau')d\tau'$
with respect to  $\tau = T/(2\pi \rho_{av})$ is provided in fig.~3.
For generic integrable systems without degeneracies in
periodic orbit lengths, $ I(\tau) = \tau$ while in the
present case, one observes a non-linear increase. 

\par For case (b), the splitting mechanism needs to be
incorporated and for this example, the ratio in which
certain orbit families split up has been arrived at by
Shudo \cite{shudo92}. As before, the sum in Eq.(\ref{eq:diag_b}) 
now refers to distinct lengths while the effective area 
$a_i(2\alpha_i -1)$ (denoted by $\overline{a}_i$)
is the sum of all areas occupied by degenerate orbit families
{\em  weighted appropriately by the phases}. Thus
$\overline{a}_i = \sum_k (-1)^{n_k} a_k$ where $a_k$ is the area
occupied by a family having length $l_i$ and which
undergoes $n_k$ reflections from the Dirichlet edge.

\par Once more, rather than the asymptotic proliferation
rate of periodic orbit families, it is the variation
of $\overline{a}_i$ with length \cite{pramana_97} which 
determines the form factor.
A plot of $I(\tau)$ for case (b) (see fig.~3) reveals a non-linear
increase having a smaller overall slope and a form
that is distinct from case (a).
Thus a change in BC from uniform
Dirichlet to non-uniform BC leads to a significant change in
the form factor.  This difference indeed shows up in the
spectral rigidity, $\Delta_3(L)$. In fig.~4, we compare 
the predictions of periodic orbit  theory  with the 
exact (numerical) values of $\Delta_3$ 
in the range $4 \leq L \leq 10$ \cite{fnote_L}
for cases (a) and (b). While the agreement for case (a) is
excellent, the predictions of periodic orbit theory capture
the overall behaviour in case (b). 
   
\par The discussion so far holds for all 
polygonal billiards where adjacent edges enclosing
an angle of the form $\pi/n_i$ have non-identical (NI) boundary conditions.
In such cases, periodic orbit families do not split up at this
angle {\em classically} though as demonstrated earlier, they can
split up {\em semiclassically}. For angles of the form $m_i\pi/n_i$ ($m_i > 1$)
however, orbit families do split up classically and traverse
different paths thereby reducing the extent   of periodic
orbit families. Thus different sets of boundary conditions
only result in an overall phase factor for each family 
and hence do not significantly affect spectral measures.
In exceptional cases however, the effect of boundary conditions
can be significant. This can be observed when 
the angle is of the form $m_i\pi/n_i$ ($m_i > 1$)
but close to an {\em integrable} 
wedge. As an example, consider the rigidity $\Delta_3 (L)$
for the irrational triangle ($\pi/2,\pi/\sqrt{9.1}$) which is
close to the integrable ($\pi/2,\pi/3)$ enclosure (fig.~5). Case (a)
clearly exhibits fluctuations close to Poisson while case (b)
shows typical GOE fluctuations for the energy range
considered. Note that the triangle has infinite genus
though over short time scales (less than the Heisenberg time)
the dynamics hovers
around its integrable counterpart
while even after $10^9$ reflections from the boundary,
parts of the constant
energy surface remain unexplored.
The two {\em non-integrable} acute angles
however serve to split up periodic orbit families
though the lengths of the resulting families remain
close to that of the original family in the
integrable enclosure.
This subtle re-organization of periodic orbit families
leads to Poisson fluctuations in case (a)
since the degeneracies in 
orbit actions which exist for the ($\pi/2,\pi/3$)
triangle get lifted in the ($\pi/2,\pi/\sqrt{9.1}$) enclosure. 
On the other hand, when the lifting of degeneracies is
accompanied by a difference in phase between two {\em split}
families (case (b)), the change is significant and leads to
GOE like fluctuations for the energy range considered
\cite{varies}.

\par 
In summary, we have demonstrated that 
a change from uniform Dirichlet to non-identical boundary
conditions on the edges of triangular billiard
can lead to significant changes in fluctuation measures.
This can be observed when the angle enclosed by the
edges with NI boundary conditions is of the form
$\pi/n$ or sufficiently close to it \cite{not_always}.
The mechanism
involved is {\em quantum splitting} due to
which adjacent families having (almost) identical
lengths acquire different phases leading to a significant
drop in contribution from both the diagonal and
off-diagonal terms in the form factor.
In particular, we have shown that it is possible to
explain the spectral fluctuations of the ($\pi/2,\pi/3,\pi/6$)
triangle when the boundary conditions are not identically
Dirichlet using periodic orbit theory.
{\em We conclude by noting that there exist quantum systems
whose density correlations cannot be determined
fully by the classical spectrum.}


\pagebreak
\centerline{FIGURE CAPTIONS}
\vskip 1.25 in

\noindent
Fig.~1.~$\Delta_3 (L)$ for the
$(\pi/2,\pi/3,\pi/6)$ triangle with (a) Dirichlet boundary
conditions on all edges ($\triangle$) (b) Neumann boundary
conditions on the edges enclosing the right angle and
Dirichlet on the third ($\diamond$). The averaging interval is 
[$\epsilon_n - \Delta\epsilon, \epsilon_n +
\Delta\epsilon$] with $\epsilon_n = 800$
and $\Delta\epsilon = 300$. The straight line and the smooth
curve are respectively the Poisson ($L/15$) and GOE results.
\vskip 0.25 in
\noindent
Fig.~2.~Two periodic orbits F1 (thick line) and F2 belonging to the same
classical family.
\vskip 0.25 in
\noindent
Fig.~3.~A plot of $I(\tau)$ for the 
$(\pi/2,\pi/3,\pi/6)$ billiard. The curve marked (integrable)
is for case (a) while (PI) represents case (b).
Also shown are three lines with slopes 1.0, 0.85 and 0.45
marked (c), (d) and (e) respectively. For averaging, see fig.~1.
\vskip 0.25 in
\noindent
Fig.~4.~The chain and dotted curves
are the exact values of $\Delta_3(L)$ for case (a) and
(b) respectively. The diamonds and squares are 
estimates obtained using periodic orbits for case (a) and (b)
respectively. For averaging, see fig.~1.
\vskip 0.25 in
\noindent
Fig.~5.~The rigidity for the irrational triangle
$(\pi/2,\pi/\sqrt{9.1})$.
Case (a) exhibits Poisson fluctuations ($\diamond$) while case (b)
shows GOE fluctuations ($+$) when $\epsilon_n = 500$ and
$\Delta\epsilon = 150$.


\begin{references}
\bibitem{fnote_rational} Irrational triangles are possibly ergodic
\cite{hob+cas}. The ones accessible to computations are
rational and hence pseudo-integrable.
\bibitem{intro_genus} The invariant surface is labeled by the genus, $g$
where $2g$ is the number of  cuts required to make the
surface singly connected.
\bibitem{as_in_pi} In generic pseudo-integrable enclosures,
$\{a_i\}$ are orbit dependent \cite{db_rapid}.
\bibitem{fnote_L} Due to limited number of orbits,
the fit is not expected to be good for small L.
\bibitem{varies} Though we
observe GOE-like fluctuations in several cases,
these quantum systems do not necessarily
belong to any universality class. There is a strong
energy dependence that manifests clearly in $\Delta_3$
but is not so significant in the nearest neighbour distribution.
\end{references}
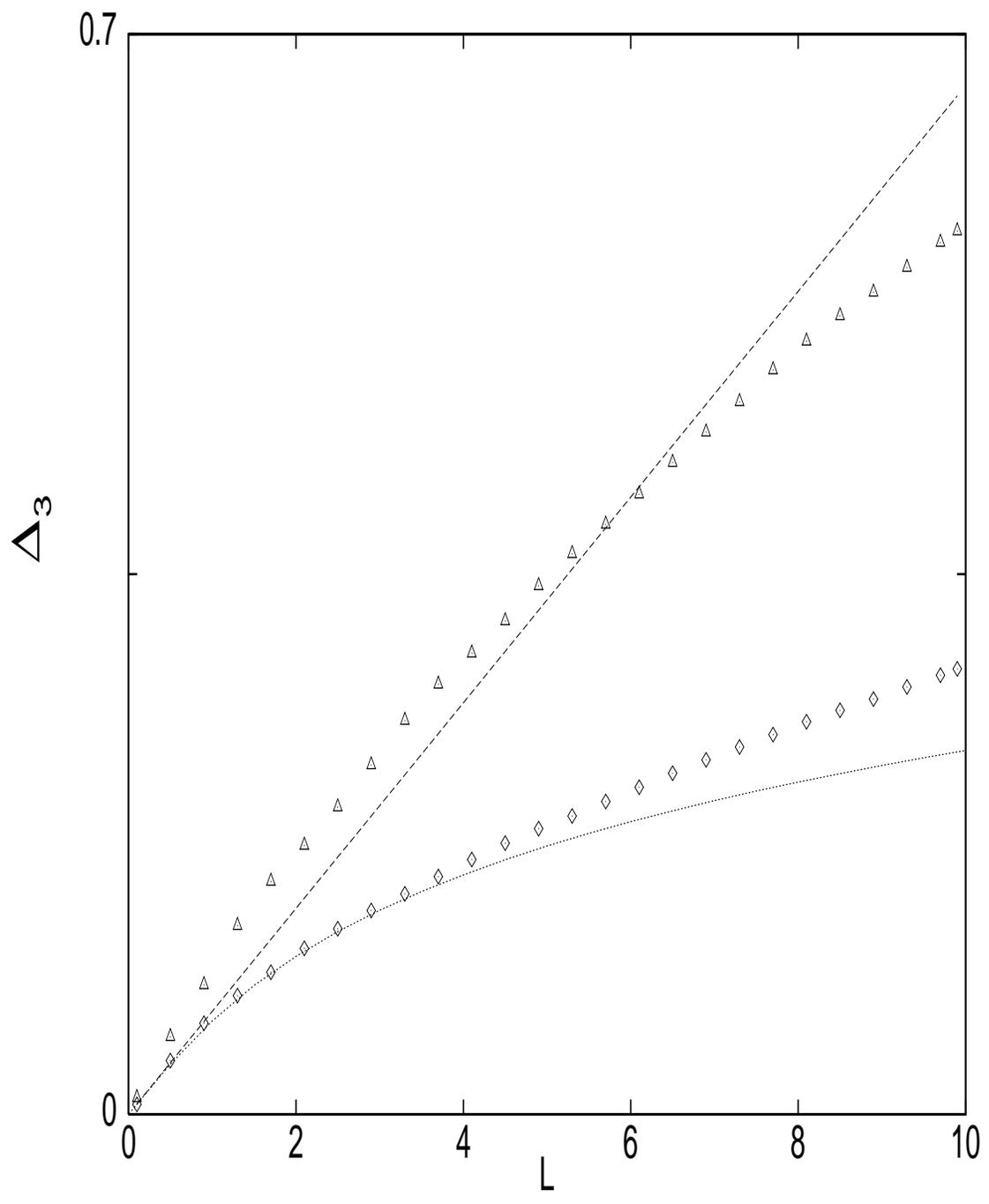
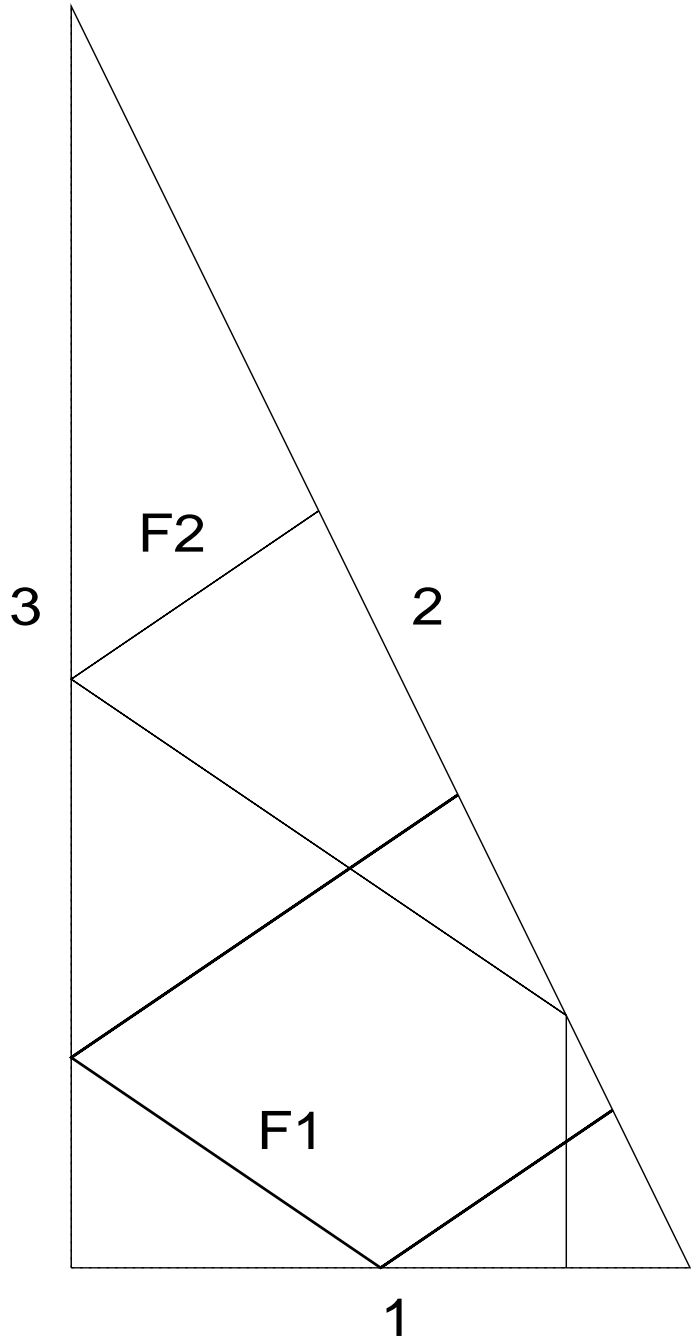
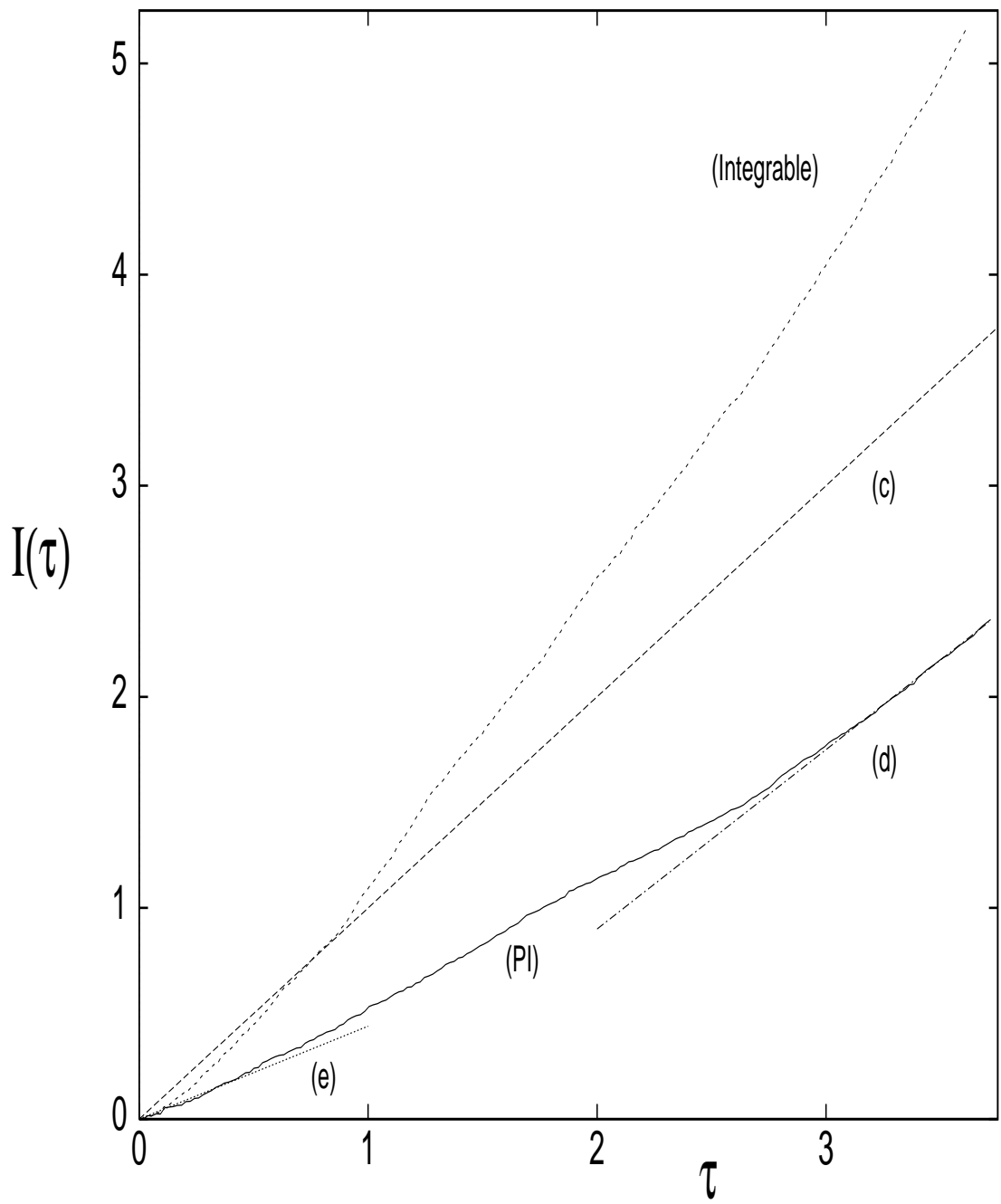
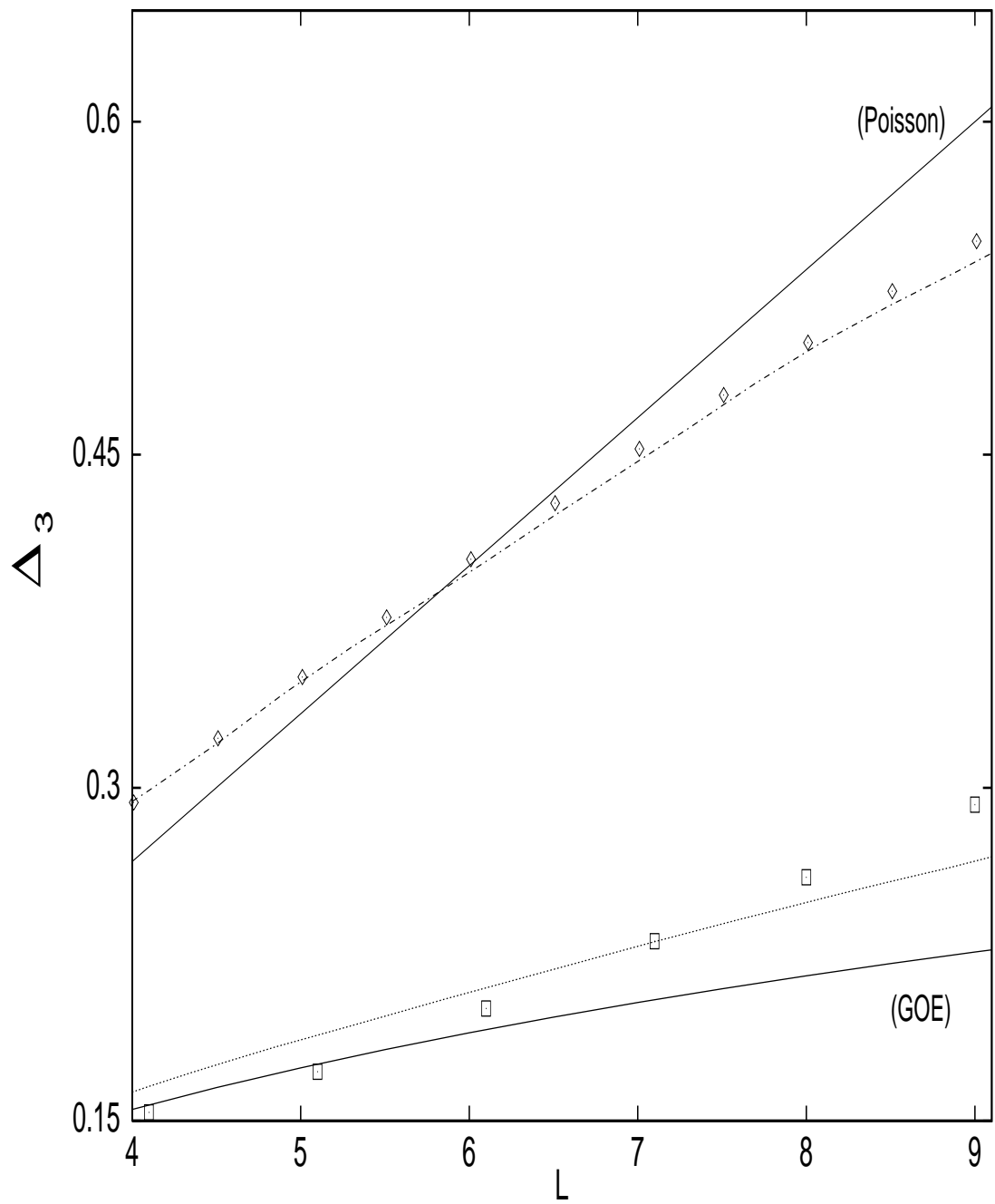
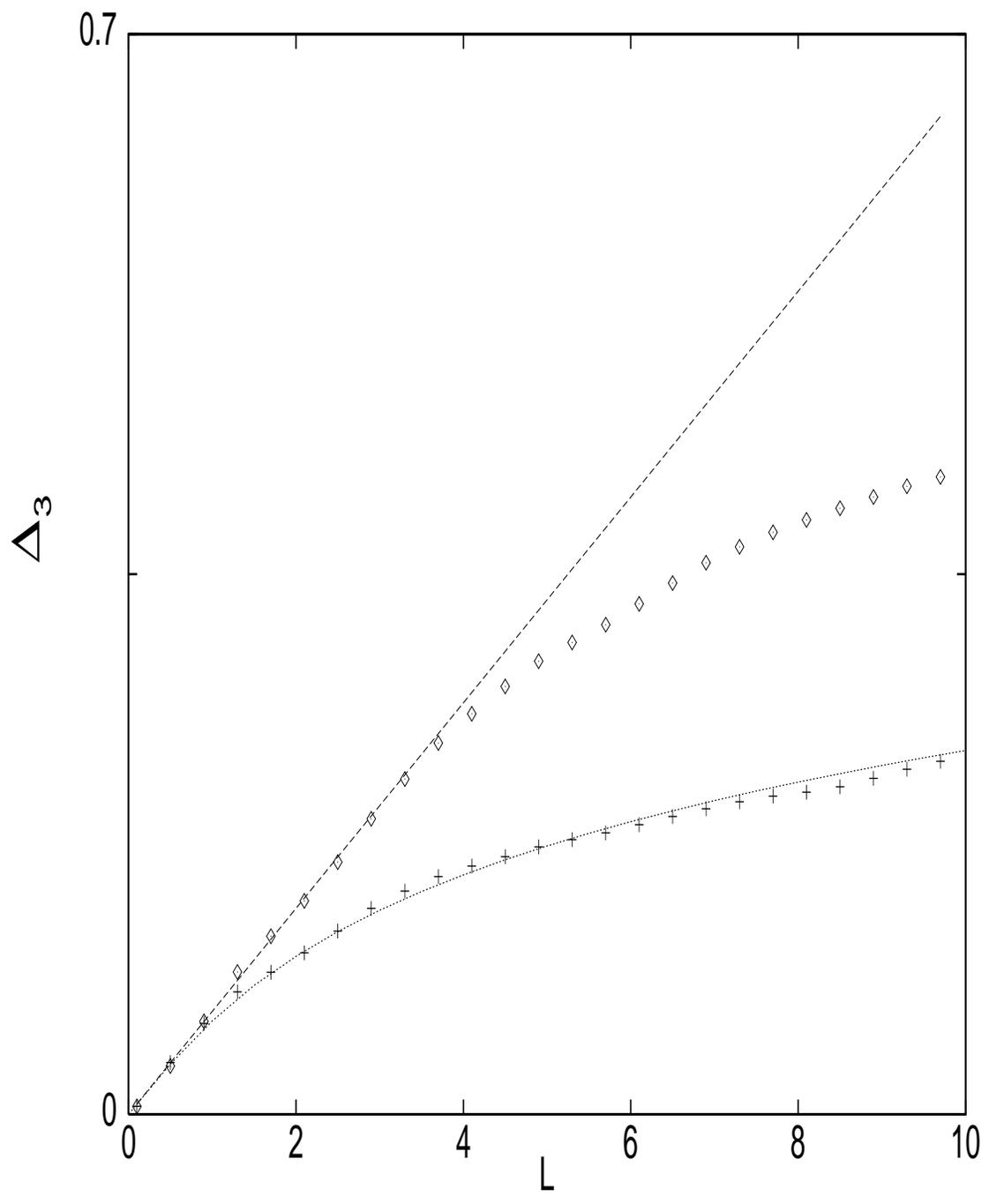
\end{document}